\documentclass[prl,twocolumn,showpacs,superscriptaddress,floatfix]{revtex4-1}
\usepackage{graphicx}
\usepackage{epstopdf} 
\usepackage{amsmath}
\usepackage{epsfig}
\usepackage{latexsym}
\usepackage{amsmath}
\usepackage{graphics}
\usepackage{bm}

\usepackage{mathtools}

\usepackage{helvet}

\begin{document}

\date{\today}
\author{J. Mumford}
\affiliation{Department of Physics and Astronomy, McMaster University,
  1280 Main St.\ W., Hamilton, ON, L8S 4M1, Canada}
\author{E. Turner}
\affiliation{Department of Physics and Astronomy, McMaster University,
  1280 Main St.\ W., Hamilton, ON, L8S 4M1, Canada}
\author{D. W. L. Sprung}
\affiliation{Department of Physics and Astronomy, McMaster University,
  1280 Main St.\ W., Hamilton, ON, L8S 4M1, Canada}
\author{D. H. J. O'Dell}
\affiliation{Department of Physics and Astronomy, McMaster University, 1280 Main St.\ W., Hamilton, ON, L8S 4M1, Canada}

\title{Quantum spin dynamics in Fock space following quenches: Caustics and vortices}

\begin{abstract}
Caustics occur widely in dynamics and take on shapes classified by catastrophe theory. At finite wavelengths they produce interference patterns containing networks of vortices (phase singularities). Here we investigate caustics in quantized fields, focusing on the collective dynamics of quantum spins. We show that, following a quench, caustics are generated in the Fock space amplitudes specifying the many-body configuration and which are accessible in experiments with cold atoms, ions or photons. The granularity of quantum fields removes all singularities, including phase singularities, converting point vortices into nonlocal vortices that annihilate in pairs as the quantization scale is increased. Furthermore, the continuous scaling laws of wave catastrophes are replaced by discrete versions. Such `quantum catastrophes' are expected to be universal dynamical features of quantized fields.
\end{abstract}

\maketitle

\begin{figure}[t]
\begin{center}
\includegraphics[width=1.0\columnwidth]{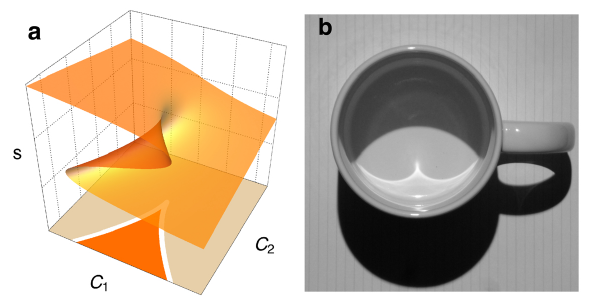}
\end{center}
	\caption{\textbf{(a)} The cusp catastrophe is given by the projection of the folds in the surface $\partial I_{c} /\partial s =0$ down onto the control plane ($C_{1},C_{2}$).   \textbf{(b)} Catastrophes are structurally stable and hence  occur generically without fine tuning or special symmetry: perfect parabolic mirrors focus light to a point but a generic curved surface like the side of a cup gives a cusp.}
	\label{fig:cuspfig}
\end{figure}

Caustics are a wave focusing phenomenon familiar as rainbow arcs \cite{nye99}, twinkling starlight \cite{berry77} and the lines of focused sunlight on the base of a cup, see Fig.\  \ref{fig:cuspfig}. They also occur in fluids as rogue waves \cite{hohmann10}, tidal bores \cite{berry18}, and large scale structure in the universe \cite{Zeldovich82,Feldbrugge18}. Defined as regions of diverging intensity in the short wavelength limit, caustics commonly take on characteristic shapes, such as the cusp. The reason for this is provided by Thom's catastrophe theory:  only singularities with these shapes are structurally stable against perturbations and hence occur universally \cite{thom75,arnold75}.  Thom found seven `catastrophes' in up to four dimensions, each forming an equivalence class with its own scaling relations analogous to the universality classes of equilibirum phase transitions \cite{berry81}. Indeed, at the heart of both caustics and phase transitions lie singularities. However, caustics also occur in non-equilibrium dynamics and in this letter we describe their morphology in dynamical \emph{quantum fields}.

In two dimensions the structurally stable catastrophe is the cusp described by a quartic generating function  $I_{c}(C_{1},C_{2};s) = C_1 s + C_2 s^2 + s^4$, where $(C_{1},C_{2})$ are control parameters (coordinates), and $s$ is a state variable. In physical applications $I_{c}$ is the \emph{action} and $s$ labels paths  \cite{berry81}. Classically allowed paths satisfy the principle of stationary action  $\partial I_{c}/\partial s = 0$ which is plotted as a surface in Fig.\ \ref{fig:cuspfig}(a). The folded portion has three solutions above each point $(C_{1},C_{2})$ whereas the non-folded portion has just one. The boundary between them forms a cusp  $C_{1} = \pm \sqrt{8/27} (-C_{2})^{3/2} $ in the control plane, which is the geometric catastrophe. It consists of two curves where the action is stationary to higher order: $\partial^2 I_{c}/\partial s^2 = 0$, giving the locus of points where two solutions coalesce. 

The corresponding wave theory with wavenumber $k$ uses $I_{c}$ to form a path integral over all paths:
\begin{equation}
\Psi_{c}(C_{1},C_{2};k)=
\sqrt{k}\int_{-\infty}^\infty  \mathrm{e}^{\mathrm{i} k( C_{1} s+ C_{2} s^2 + s^4 )} \mathrm{d}s .  \label{eq:pearcey}
\end{equation}
For $k=1$ this is the Pearcey function $\mathrm{Pe}(C_{1},C_{2})$ which is the universal `wave catastrophe' dressing a cusp \cite{pearcey46,handbook}.  It is straightforward to show that $\Psi_{c}=k^{\beta}$ $ \times \mathrm{Pe}(k^{\sigma_1}C_{1},k^{\sigma_2}C_{2})$ where $\beta=1/4$ is the Arnold index governing the scaling of the amplitude with $k$ and the exponents  $\sigma_1 = 3/4$ and  $\sigma_2 = 1/2$ are Berry indices that govern the fringe spacings in the $(C_{1},C_{2})$-plane  \cite{berry77}.  Each class of catastrophe has its own set of scaling exponents but the same general morphology: at large scales ($k \rightarrow \infty$) we retrieve geometric caustics with diverging amplitude, but at wavelength scales interference removes the divergences to produce smooth oscillatory patterns. At the finest scales these patterns contain a network of vortex-antivortex pairs \cite{pearcey46,handbook}.  Plots of $\mathrm{Pe}(C_{1},C_{2})$ are given in the Supplementary Material (SM) \cite{SM}.

 `Quantum catastrophes' are an extension of these ideas: they occur when wave theory itself is singular and we must (2nd-) quantize the field in order to regulate it.  Leonhardt has given the example of the logarithmic phase singularity suffered by a wave crossing an event horizon and argued that it is resolved in quantum field theory by the emission of photons as Hawking radiation \cite{leonhardt02}; Berry and Dennis considered optical phase dislocations (vortex lines) where the phase is undefined and emphasized the role played by vacuum fluctuations \cite{berry04,berry08}. 

Many-body dynamics provides another stage for exploring quantum catastrophes \cite{odell12,Mumford17}.  Consider the 
transverse-field Ising model (TFIM) for $N$ spins
\begin{equation}
\hat{H}_{\mathrm{TFIM}}= \frac{1}{N} \sum_{i < j} J_{ij}  \ \hat{\sigma}_{i}^{(z)} \hat{\sigma}_{j}^{(z)} - \Omega \sum_{i} \hat{\sigma}_{i}^{(x)}
\end{equation}
where the $\hat{\sigma}$'s are Pauli operators,  and $J_{ij}$ and $\Omega$ control spin-spin interactions and the transverse field, respectively. The TFIM can be simulated using trapped ions where two internal states act as spin states $\vert \! \uparrow \rangle$ and $\vert  \! \downarrow \rangle$,  and $J_{ij}=J/\vert \mathbf{r}_{i} -\mathbf{r}_{j}\vert^{\epsilon}$ is engineered by coupling motional and spin degrees of freedom using lasers  \cite{Britton12,Jurcevic14,Richerme14}. In Ref.\  \cite{Bohnet16} several hundred ions were prepared in a $\hat{\sigma}^{(x)}$ eigenstate following which $J_{ij}$ was switched on, generating spin entanglement. In these experiments $\epsilon$ was as small as 0.02 so that $J_{ij}$  becomes independent of position and the system reduces to a two-mode (spin up/down) quantum field described by the Hamiltonian \cite{Das06}
\begin{equation}
\hat{H}_{2M}=\frac{2J}{N} \hat{S}_{z}^2-2 \Omega \hat{S}_{x}
\end{equation}
where $\hat{S}_z=\sum_{i}^{N} \hat{\sigma}_{i}^{(z)}/2$ etc.\ are collective spin operators.
Bose-Einstein condensates (BECs) forming Josephson junctions also realize $\hat{H}_{2M}$  using either two internal states or by trapping atoms in a double well potential \cite{Milburn97,albiez05,Levy07,zibold10,Gerving12,Valtolina15,trenkwalder16}. The two polarization states of optical beams provide another example of a two-mode system; nonlinearity can be added in a Kerr medium and configured so as to give polarization squeezing \cite{Rigas13,luis02,korolkova05}.  

States evolving under $\hat{H}_{2M}$ live on a generalized Bloch sphere described by the vector  $\hat{\mathbf{S}}=(\hat{S}_x,\hat{S}_y,\hat{S}_z)$ of length $N/2$ \cite{Bohnet16,zibold10}. We work in the  $z$-basis satisfying  $\hat{S}_{z} \vert N/2, m \rangle = m \vert N/2, m \rangle$, where $m = \! {\textstyle \frac{1}{2}} \times$ (No.\ of $\uparrow$ spins $-$ No.\ of $\downarrow$ spins). Defining  $z \equiv 2 m /N$, which takes values between -1 and +1 in steps of $2/N$, we henceforth denote  $ \vert N/2, m \rangle$ by $\vert z \rangle$ and a general state is written 
\begin{equation}
\vert \Psi (t) \rangle = \sum_{z} a_{z}(t) \vert z \rangle
\end{equation}
where $a_{z}(t)$ are Fock-space amplitudes.  The conjugate variable to $z$ is the phase difference $\phi \equiv \phi_{\uparrow} - \phi_{\downarrow}$ between the two modes. Defining a quantum operator $\hat{\phi}$ is problematic, but in the semiclassical regime it can be argued that $[ \hat{\phi}, \hat{z}] \approx 2 \mathrm{i}/N $ \cite{Leggett,Nieto93}, where $2/N$ is analogous to $\hbar$ in single-particle quantum mechanics.

 \begin{figure}[!t]
\begin{center}
\includegraphics[width=1.0\columnwidth]{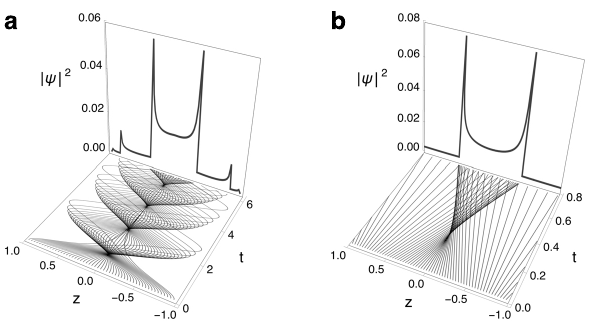}
\end{center}
	\caption{Geometric cusps in Fock space as a function of time (scaled as $t \rightarrow \Omega t/\hbar$) in a two mode field.   The initial state has a phase difference $\phi=0$ and $\Lambda$ is quenched at $t=0$.  \textbf{(a)} Rays generated by the classical field Hamiltonian $H_{\mathrm{CF}}$ form a train of cusps.  \textbf{(b)}  The kicked version $H_{\mathrm{kick}}$ generates a single cusp starting at $t_{0}=1/4\Lambda$. The rays are now straight:  $z(t)= - 2 t \sin ( 2\Lambda z_0)+z_0$ [the phase rays are constants: $\phi(t)=\phi_{0}= 2 \Lambda z_{0}$] but structural stability means the singularity is still a cusp, where the control parameters are $C_{1} = z (3/2t)^{1/4}$ and $C_{2} = (\frac{1}{4 \Lambda} -t)\sqrt{6/t}$.
In both images $\Lambda = 2.1$ and the back panels show the probability density obtained by binning the rays at the final time shown.} 
	\label{fig:trajfig}
\end{figure}

\emph{Ray caustics}---In the classical field (CF) limit $N\rightarrow \infty$, and  $(z,\phi)$ are continuous commuting variables.   Quantum fluctuations can be mimicked to some degree via the truncated Wigner approximation   \cite{Polkovnikov10,Javan13}  where multiple initial conditions are sampled from a quantum distribution but propagated using the classical equations  $\dot{\phi} = \partial_{z} H_{\mathrm{CF}}$ and $\dot{z}=- \partial_{\phi} H_{\mathrm{CF}} $, where $H_{\mathrm{CF}} \equiv \lim_{N \to \infty}H_{2M}/N\Omega = \Lambda z^2/2 - \sqrt{1-z^2} \cos \phi$ \cite{smerzi97}, and $\Lambda = J/\Omega$. Fig.\ \ref{fig:trajfig} shows this approach applied to a quench where $\Lambda$ is changed from $0$ to $2.1$ at $t=0$. Each `ray' $z(t)$ has a different initial number difference $z_{0}$: for the initial quantum distribution we choose a completely undefined $z_{0}$ implying a well defined phase difference. This corresponds to a paramagnetic state ($\hat{\sigma}^{(x)}$ eigenstate). Thus, the set of initial points $\{z_{0}\}$ is uniformly distributed over the range $-1 \le z_{0} < 1$ and when propagated with a finite value of $\Lambda$ the envelopes of the rays produce a train of cusp shaped caustics as shown in Fig.\ \ref{fig:trajfig}(a).  These ray caustics give the geometric level of catastrophe and cause  divergences in the probability density as seen on the back panel.  

We now switch to a kicked Hamiltonian where $J$ is flashed on and off once. As shown in Fig.\ \ref{fig:trajfig}(b), this gives a single cusp which avoids interference with subsequent cusps and allows us to perform a quantum calculation analytically.  As we are interested in the generic part of the cusp near $z=0$, rather than the deformed part near the edges at $\vert z \vert =1$, we replace $\sqrt{1-z^2} \to 1$ (valid for times before the cusp reaches $\vert z \vert =1$ \cite{SM}), giving $H_{\mathrm{kick}}/N \Omega =  \Lambda \delta(t) z^2 /2 -  \cos \phi$.

The appearance of cusp caustics in 2D ($z+ \mathrm{time}$) is generic.
 We find similar caustics for other choices of parameters and initial conditions, including the opposite quench where $\Lambda$ goes from  large to  small.   The cusp train corresponds to quantum revivals \cite{Milburn97,Veksler15}; related structures occur in the dynamical diffraction of light \cite{Berry66},  kicked rotors \cite{Leibscher04}, and BECs in optical lattices \cite{Huckans09}.

\begin{figure*}[!h]
\centering
\includegraphics[width=5in]{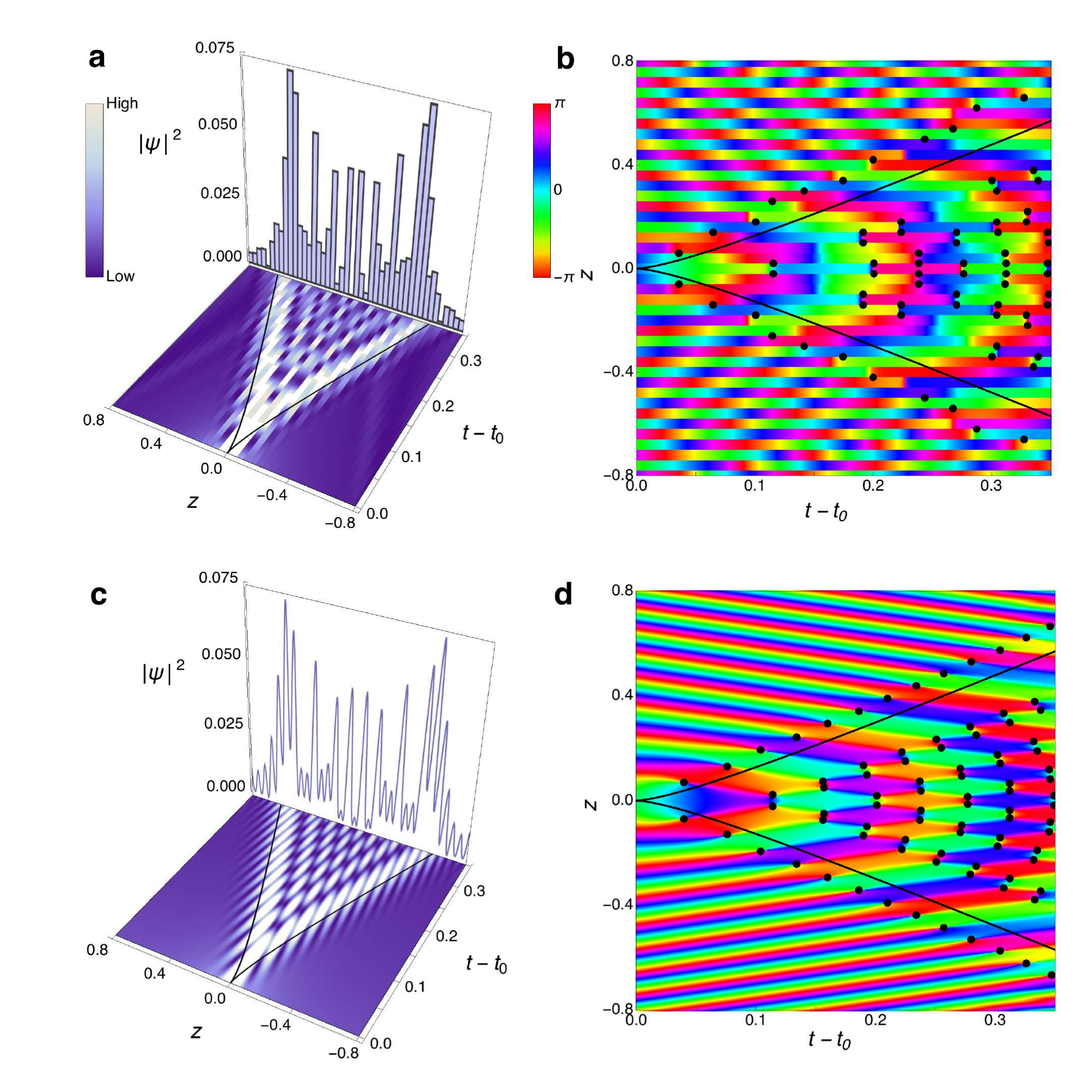}
	\caption{A quantum catastrophe (upper row) versus a wave catastrophe (lower row). The former is fundamentally discrete, the latter continuous. In both cases divergences present in the geometric cusp (black curves) in Fig.\ \ref{fig:trajfig} are removed [unlike Fig.\ \ref{fig:trajfig}, we have only plotted times $t>t_{0}$, where $t_{0}=1/(4 \Lambda)$ is the time when the geometric cusp starts].  Parts \textbf{(a)} and \textbf{(b)} plot
the probabilities $\vert a_{z}(t)\vert^2$, and phases $\mathrm{Arg}[ a_{z}(t)]$, respectively, of the Fock amplitudes. These are obtained using a numerical solution of the quantum dynamics [see Eq.\ (SM7)]  and live on a discrete grid which we represent graphically as ribbons of finite width $2/N$.  Parts \textbf{(c)} and  \textbf{(d)} plot $\Psi_{\mathrm{CA}}$ given in Eq.\ (\ref{eq:wf2}). Black dots mark phase dislocations: there is a single row of vortices on each side outside the cusp and a pattern of counter-rotating vortex-antivortex pairs inside. However, in the quantum catastrophe all phase singularities are regulated by granularity with some dislocation pairs removed altogether. In this figure $N = 50$ and the vortices in each pair are separated by a single grid spacing, but when $N>10^4$ we find some pairs separated by two grid spacings. As explained in the main text, the number of grid spacings within a vortex pair is expected to grow as $\mathcal{R} \sim N^{1/4}$.   The back panels in (a) and (c) plot the probability density at the final time shown. In all images $\Lambda = 2.1$, $\alpha=0.53$.  } 
	\label{fig:wfpanel}
\end{figure*}

\emph{Quantum caustics}---The divergences present in the CF caustic are cured by quantizing the field \cite{odell12}. The resulting quantum catastrophe is displayed in the upper row of Fig.\ \ref{fig:wfpanel}. It is formed of the set of amplitudes $\{ a_{z} (t) \}$  which we find by solving the many-particle Schr\"{o}dinger equation $\mathrm{i} \hbar \partial_{t} \vert \Psi \rangle = (\hat{H}_{2M})_{\mathrm{kick}} \vert \Psi \rangle$ numerically; for details see the SM \cite{SM} (the initial state has been taken to be a Gaussian in Fock space $ \vert \Psi_0 \rangle =  \alpha^{-1/2} (2/\pi)^{1/4} \sum_{z} \mathrm{e}^{- (z/2 \alpha)^2} \vert z \rangle$ with width $\alpha$).

In the semiclassical ($N \gg 1$)   regime we can proceed analytically. Employing the time evolution operator $\mathcal{\hat{U}}(t,t_{i}) = \mathcal{T} \{ \exp [-(\mathrm{i}/\hbar) \int_{t_{i}}^{t} \hat{H}_{\mathrm{kick}}(t') dt' ] \} = \exp [\mathrm{i} (t-t_{i}) N \cos \hat{\phi}] \, \exp [-\mathrm{i} \Lambda N \hat{z}^2/2]$,   where $\mathcal{T}$ is the time ordering operator  and $\hat{\phi}$ is the phase operator \cite{haake09}, we have $\vert \Psi (t) \rangle = \mathcal{\hat{U}}(t,t_{i}) \vert \Psi_{0} \rangle $.
 Both $\hat{\phi}$ and $\hat{z}$  have discrete spectra when $N$ is finite and their eigenfunctions form a discrete Fourier transform pair \cite{pegg89,SM}. We obtain
\begin{eqnarray}
&& \Psi(z,t;N)      =      \langle z\vert \mathcal{\hat{U}}(t) \vert \Psi_0
\rangle   =  \frac{(2/\pi)^{1/4}}{\alpha^{1/2} (N+1)}   \nonumber \\
 & & \times    \sum_{p,m = -N/2}^{N/2} \mathrm{e}^{-(\frac{z_{m}}{2 \alpha})^2}
   \mathrm{e}^{-\mathrm{i} N (
      \frac{\Lambda  z_{m}^2}{2} -  t \cos \phi_p - \frac{z_{n}-z_{m}}{2} \phi_p  )}  \quad
\label{eq:wf1}
\end{eqnarray}
where $\phi_p =\frac{2 \pi p}{N + 1}$ and $z_{m}=\frac{2m}{N}$. Poisson resummation of the $m$-sum (exact when limits are $\pm \infty$) gives \cite{SM,Poisson}
\begin{equation}
\Psi \approx \frac{(\pi/2)^{\frac{1}{4}}}{(\alpha N \Lambda)^{\frac{1}{2}}}  \sum_{k=-\infty}^{+\infty}   e^{-(\frac{u_{k}}{4 \Lambda \alpha})^2}  e^{\mathrm{i}
  N \left (\frac{u_{k}^2}{8 \Lambda} + t \cos u_{k} +\frac{z u_{k}}{2} \right)}  
\label{eq:discretewf2}
\end{equation}
where $u_{k} =2 \pi k /(N+1)$.  To obtain the  wave catastrophe plotted in the lower row of Fig.\ \ref{fig:wfpanel} we apply the continuum approximation (CA) where we let $\phi$ and $z$ become continuous. This changes Eq.\ (\ref{eq:discretewf2}) into an integral: 
\begin{equation}
\Psi_{\mathrm{CA}} \sim  \frac{(N/\alpha \Lambda)^{\frac{1}{2}}}{(32 \pi^3)^{\frac{1}{4}} }   \int_{-\infty}^{+ \infty}  e^{-(\frac{u}{4 \Lambda \alpha})^2}  e^{\mathrm{i}
  N \left (\frac{u^2}{8 \Lambda} + t
    \cos u +\frac{z u}{2} \right) }    \mathrm{d}u  .
\label{eq:wf2}
\end{equation}

\emph{Quantum Pearcey function}---The above analysis suggests the existence of a universal discrete counterpart to Eq.\ (\ref{eq:pearcey}).  When $N\gg 1$ the dominant contributions to the sum in Eq.\ (\ref{eq:discretewf2}) come from the neighborhoods of stationary points; we can capture these by expanding $\cos u$ up to $u^4$. Defining the variable $s = (Nt/24)^{1/4} u$  yields
\begin{equation}
\Psi_{\mathrm{Qu.}\mathrm{Prcy.}} =  \frac{(\pi/2)^{\frac{1}{4}}}{(\alpha N \Lambda)^{\frac{1}{2}}} \sum_{s=-\infty}^{\infty}  e^{-\frac{\zeta s^2}{\sqrt{N}}} \, e^{\mathrm{i}
   \left (N^{\frac{3}{4}}C_{1} s +  N^{\frac{1}{2}}C_{2} s^2 +s^4 \right)}  
\label{eq:discretewf3}
\end{equation}
where $C_{1}(z,t) = z (3/2t)^{1/4}$, $C_{2}(t) = (\frac{1}{4 \Lambda} -t)\sqrt{6/t}$, and $\zeta(t) =
\sqrt{\frac{3}{2 t}} \frac{1}{4 \Lambda^2 \alpha^2}$. Up to an innocuous gaussian envelope inherited from $ \Psi_{0}$, Eq.\ (\ref{eq:discretewf3}) is a discrete Pearcey function with control parameters $(N^{\frac{3}{4}}C_{1}, N^{\frac{1}{2}}C_{2} )$ where $N$, $s$ and $C_{1}$ take on discrete values.  To retrieve the continuous Pearcey function we can make the same approximations as we used to go from Eq.\ (\ref{eq:discretewf2}) to (\ref{eq:wf2}). Letting $\alpha \gg 1$ so that the gaussian can be dropped, we finally obtain  
\begin{equation}
\Psi_{\mathrm{CA}} \sim (3 N/4t \pi^3 (\alpha \Lambda)^2)^{\frac{1}{4}}  \, \mathrm{Pe}(N^{\frac{3}{4}} C_{1},N^{\frac{1}{2}} C_{2}).
\label{eq:wf3}
\end{equation}

\emph{Vortices}---Wave catastrophes contain networks of vortices which are \emph{points} where the phases of the amplitudes $a_{z}(t)= \rho (z,t) \mathrm{e}^{\mathrm{i} \theta(z,t)}$ ($\theta$ should not be confused with $\phi$) take all values  and hence are undefined (phase singularities) and $\rho=0$. These are found by evaluating the phase change  $\Delta \theta=\oint \nabla \theta \cdot \mathrm{d} \mathbf{l}$ around all possible circuits. Each singly charged vortex gives $\Delta \theta= \pm 2 \pi$. 

\begin{figure}[t!]
\centering
\includegraphics[width=1.0\columnwidth]{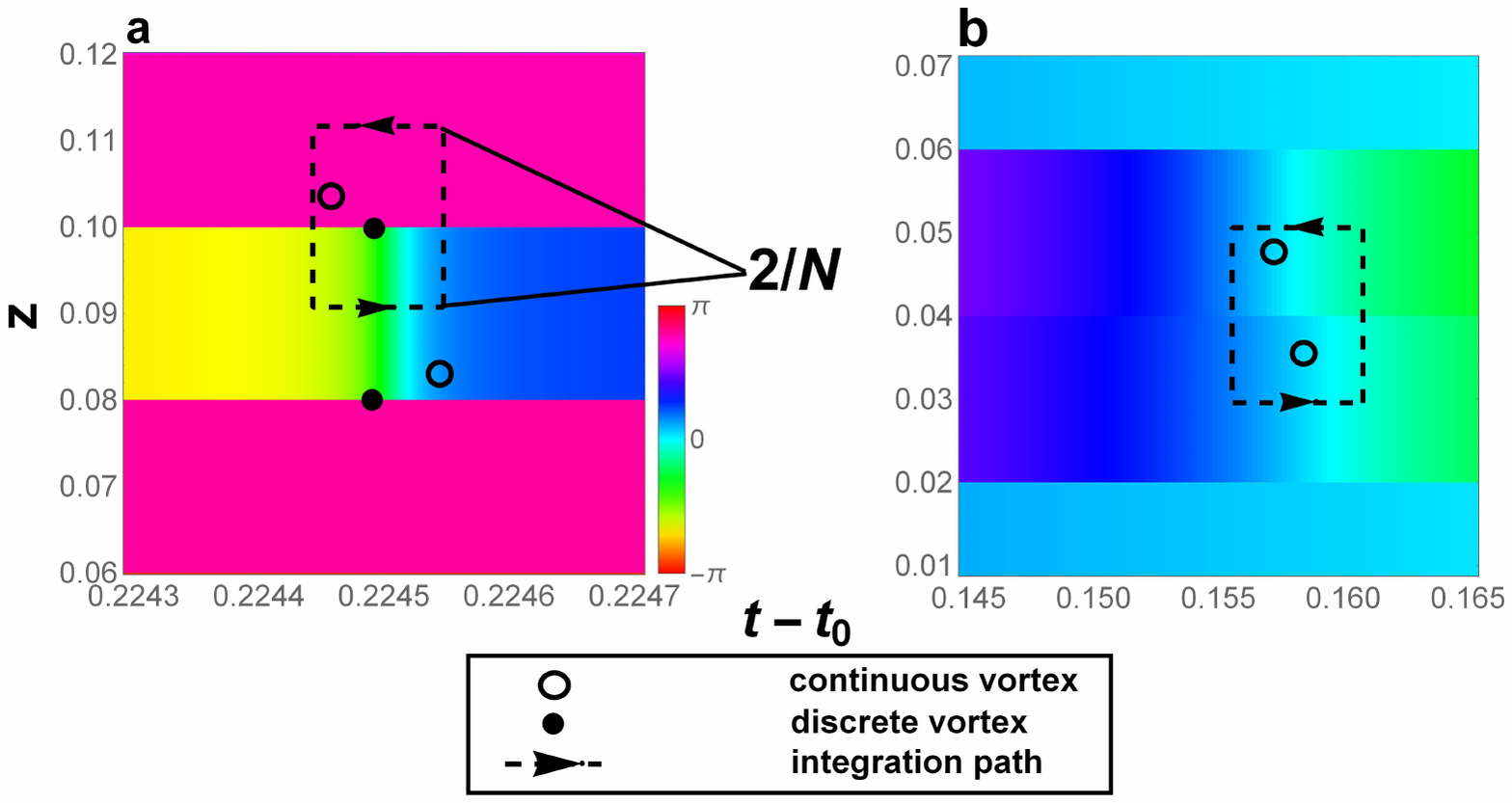}
	\caption{Two magnified regions of Fig.\ \ref{fig:wfpanel}(b) illustrating the morphology of vortices in a quantum catastrophe.   \textbf{(a)} A vortex-antivortex pair (filled circles) that survive quantization, together with the location of the same pair in the continuum approximation (unfilled circles) from Fig.\ \ref{fig:wfpanel}(d). Dashed lines indicate an integration circuit exaggerated in the time (horizontal) direction to make it visible. When just one vortex is enclosed by the integration path it survives discretization.  \textbf{(b)} When both members of a pair are enclosed they annihilate. The phase changes rapidly near a vortex and hence the horizontal scale of panel (a) is more magnified than in (b).  }
	\label{fig:annihilvort}
\end{figure}

Remarkably, we find that computing the integral for the exact quantum case (with a `minimum phase difference' rule for handling the discrete steps in the circuit--see \cite{SM}) also yields circuits where $\Delta \theta=\pm 2 \pi$. However, the discretization of Fock space prevents true phase singularities: the $a_{z}(t)$ live on a grid and are always single valued. Thus, singular points are replaced by nonlocal vortices \cite{Desyatnikov11} which are distributed over two or more sites where the phase on each site is always well defined, see Fig.\ \ref{fig:annihilvort}. The dots in Figs.\ \ref{fig:wfpanel}(b) and \ref{fig:annihilvort}(a) therefore indicate non-vanishing circuit integrals rather than  phase singularities (their exact locations along $z$ are ambiguous: we  plot them between the two sites that share the vortex).

One consequence of granularity is that some Pearcey vortices are missing:  we expect a pair to annihilate when they fall within the same integration circuit, see  Fig.\ \ref{fig:annihilvort}.   According to Eq.\ (\ref{eq:wf3}), the wave catastrophe is proportional to $\mathrm{Pe}(z [3N^{3}/2t]^{1/4}, (6N/t)^{1/2}[\frac{1}{4 \Lambda}-t])$. Focusing on the dependence along $z$, the distance between any two points, and in particular between the two members of a vortex pair, scales as  $d_\mathrm{v} \propto N^{-3/4}$. Forming the ratio with the quantization length $d_{q}=2/N$ gives a resolution parameter  
$ \mathcal{R} = d_{\mathrm{v}}/d_{q} \propto N^{1/4}$. More Pearcey vortices survive when $\mathcal{R}$ is large.

\emph{Discrete scaling}---The wave catastrophe in Eq.\ (\ref{eq:wf3})  has continuous self-similar scaling  (with exponents $\beta=1/4,\sigma_{1}=3/4,\sigma_{2}=1/2$)
so that varying $N$ is equivalent to scaling the coordinates and amplitude (this has no effect on the ray caustic which is independent of $N$). We have verified numerically that the quantum catastrophe also obeys these scaling relations when $N \gg 1$ \cite{SM}.  However, at smaller $N$ the granularity becomes evident: continuous scaling is replaced by a discrete version determined by $m'/m=(t'N'/tN)^{1/4}$ and $(t'-t_{0}) \sqrt{N'/t'}=(t-t_{0})\sqrt{N/t}$ which must be simultaneously satisfied. Here $t_{0}=1/4 \Lambda$ marks the cusp tip, and $m$ and $m'$ are integers specifying $z_{m}=2 m/N$ and $z_{m'}=2 m'/N$. This is reminiscent of a quantum anomaly where a continuous scaling symmetry of the classical action becomes discrete due to quantum effects \cite{Fujikawa79,Esteve02,Coon02}. Quantum anomalies are associated with corrections to commutators \cite{Olshanii10}; whereas the wave catastrophe is consistent with the  commutator $[ \hat{\phi}, \hat{z} ] = 2 \mathrm{i} /N$ (which is distinct from a CF where $[ \phi, z ] = 0$), field quantization leads to $[ \hat{\phi}, \hat{z} ] = (2 \mathrm{i} /N)\{ \mathbf{I}  - (N+1) \vert \phi_{-N/2} \rangle \langle \phi_{-N/2} \vert \} $ \cite{SM,pegg89,Heisenbergalgebra}.  

\emph{Experimental observation}---At any time $t$ the interference pattern in Fig.\ \ref{fig:wfpanel}(a) is the probability distribution for the projection of the total spin along $z$. For atomic spins this can be measured by selectively addressing the $\vert \! \! \uparrow \rangle$ state with a laser and recording the fluorescence \cite{Bohnet16} (for atoms in a double well absorption imaging can be used to count the number in each well \cite{albiez05}). Single spin resolution (revealing quantization) can be obtained with ions which can be addressed and read out individually with an error $< 10^{-3}$ \cite{Harty14}. Resolving the phase dislocations in Fig.\ \ref{fig:wfpanel}(b) is more challenging; different spin components can be measured by first using a laser to rotate the spins before readout, but luckily we do not require full state tomography of $4^N$ measurements. In our $\frac{N}{2}$ subspace a more modest $\mathcal{O}(N^2)$ measurements suffices \cite{SM}. 

\emph{Conclusion}---Quenches in two-mode quantum fields generically give rise to cusp catastrophes in Fock space described by discrete Pearcey functions featuring nonlocal vortices. In the continuum limit these are characterized by three scaling exponents, an example of universality in many-body dynamics \cite{nicklas15}.  The cusp is a member of a hierarchy: higher-mode fields will display higher catastrophes. For the full many-body wave function these will rapidly become intractable, but in \cite{Kirkby17} we have initiated the study of catastrophes in correlation functions where the number of dimensions is greatly reduced and even simple catastrophes become relevant.

\acknowledgments We are grateful for discussions with M. Olchanyi and R. Plestid on quantum anomalies, with M. R. Dennis on discrete vortices, and with N. Akerman, T. Manovitz, and R. Shaniv on trapped ions. Funding was provided by NSERC (Canada).


\begin{thebibliography}{8}

\bibitem{nye99}{J. F. Nye,  \textit{Natural focusing and the fine
      structure of light} (Institute of Physics, Philadelphia, 1999).}

\bibitem{berry77}{M. V. Berry, Focusing and twinkling: critical
    exponents from catastrophes in non-Gaussian random short waves. J. Phys. A: Math. Gen. \textbf{10}, 2061 (1977).} 

\bibitem{hohmann10}{R. H\"{o}hmann, U. Kuhl, H.-J.
    St\"{o}ckmann, L. Kaplan,  and E. J. Heller,  Freak waves in
    the linear regime: a microwave study. Phys. Rev. Lett. \textbf{104}, 093901 (2010).}      
    
\bibitem{berry18}{M. V. Berry, Minimal analytical model for undular tidal bore profile; quantum and Hawking effect analogies. New J. Phys. \textbf{20}, 053066 (2018).}     

\bibitem{Zeldovich82}{V. I. Arnold, S. F. Shandarin,  and Ya. B. Zeldovich,  The Large Scale Structure of the
    Universe I. General Properties. One- and Two-Dimensional Models. Geophys. Astrophys. Fluid Dynamics \textbf{20}, 111 (1982).}
    
\bibitem{Feldbrugge18}{J. Feldbrugge, R. van de Weygaert, J. Hidding and J. Feldbrugge, Caustic Skeleton \& Cosmic Web},
J. Cosmol. Astropart. Phys., \textbf{2018}, 27 (2018).  

\bibitem{thom75}{R. Thom, \textit{Structural Stability and
      Morphogenesis} (Benjamin, Reading MA, 1975).}

\bibitem{arnold75}{V. I. Arnol'd, Critical points of smooth functions
    and their normal forms \textit{Russ. Math. Survs.} \textbf{30}, 1
    (1975).}

\bibitem{berry81}{M. Berry, \textit{Singularities in Waves and Rays} in Les Houches, Session XXXV, 1980 \textit{Physics of Defects}, edited by R. Balian et al. (North-Holland Publishing, Amsterdam,
1981).}   

\bibitem{pearcey46}{T. Pearcey, The structure of an electromagnetic
    field in the neighborhood of a caustic. Phil. Mag. \textbf{37}, 311 (1946).} 
    
 \bibitem{handbook}{\textit{NIST Handbook of Mathematical Functions}, edited by
Olver et al. (Cambridge University, New York, 2010), chapter 36. Available online at dlmf.nist.gov}
   
 \bibitem{SM}{See Supplemental Material at [URL will be inserted by publisher] for background information on wave catastrophes and details of both the analytical and numerical methods used in this letter, and which includes references \cite{Krahn09,Susskind64,Carruthers68,Diracbook,Berry99b,Lanyon17}.}  

\bibitem{Krahn09}{G. Krahn and D.H.J. O'Dell, Classical versus quantum dynamics of the atomic Josephson junction, J. Phys. B: At. Mol. Opt. Phys. \textbf{42}, 205501, (2009).}

\bibitem{Susskind64}{L. Susskind and J. Glogower, Quantum mechanical phase and time operator. Physics \textbf{1}, 49 (1964).}

\bibitem{Carruthers68}{P. Carruthers and M. M. Nieto, Phase and Angle Variables in Quantum Mechanics. Rev. Mod. Phys. \textbf{40}, 411 (1968).}

\bibitem{Diracbook}{P. A. M. Dirac, \textit{Principles of Quantum Mechanics}, 2nd Ed. (Clarendon, Oxford, 1936).} 

\bibitem{Berry99b}{ M. V. Berry and E. Bodenschatz, Caustics, multiply reconstructed by Talbot interference.   J. Mod. Opt. \textbf{46}, 349 (1999).  }

\bibitem{Lanyon17}{B. P. Lanyon, C. Maier, M. Holz\"{a}pfel, T. Baumgratz, C. Hempel, P. Jurcevic, I. Dhand, A. S. Buyskikh, A. J. Daley, M. Cramer, M. B. Plenio, R. Blatt and C. F. Roos, Efficient tomography of a quantum many-body system, Nat. Phys. \textbf{13}, 1158 (2017).}    

\bibitem{leonhardt02}{U. Leonhardt,  A laboratory analogue of the event
    horizon using slow light in an atomic medium. Nature \textbf{415}, 406 (2002).}

\bibitem{berry04}{M. V. Berry and M. R. Dennis,  Quantum cores of
    optical phase singularities. J. Opt. A: Pure
    Appl. Opt. \textbf{6}, S178 (2004).}

\bibitem{berry08}{M. V.  Berry, Three quantum obsessions. Nonlinearity \textbf{21}, T19 (2008).}

\bibitem{odell12}{D. H. J. O'Dell,  Quantum catastrophes and ergodicity
    in the dynamics of bosonic Josephson
    junctions. Phys. Rev. Lett. \textbf{109}, 150406 (2012).}
    
\bibitem{Mumford17}{J. Mumford, W. Kirkby, and D.H.J. O'Dell, Catastrophes in non-equilibrium many-particle wave functions: universality and critical scaling, J. Phys. B: At. Mol. Opt. Phys. \textbf{50}, 044005 (2017).}    


\bibitem{Britton12}{J. W. Britton, B. C. Sawyer, A. C. Keith, C.-C. J. Wang, J. K. Freericks, H. Uys, M. J. Biercuk, and J. J. Bollinger, Engineered two-dimensional Ising interactions in a trapped-ion quantum simulator with hundreds of spins, Nature \textbf{484}, 489 (2012).}

\bibitem{Jurcevic14}{P. Jurcevic, B. P. Lanyon, P. Hauke, C. Hempel, P. Zoller, R. Blatt, and C. F. Roos, Quasiparticle engineering and entanglement propagation in a quantum many-body system, Nature \textbf{511}, 202 (2014).}

\bibitem{Richerme14}{P. Richerme,  Z.-X. Gong, A. Lee, C. Senko, J. Smith, M. Foss-Feig, S. Michalakis, A. V. Gorshkov, and C. Monroe, Non-local propagation of correlations in quantum systems with long-range interactions, Nature \textbf{511}, 198 (2014).}

\bibitem{Bohnet16}{J. G. Bohnet, B. C. Sawyer, J. W. Britton, M. L. Wall, A. M. Rey, M. Foss-Feig, and J. J. Bollinger, Quantum spin dynamics and entanglement generation with hundreds of trapped ions, Science \textbf{352}, 1297 (2016).}



\bibitem{Das06}{A. Das, K. Sengupta, D. Sen, and
    B. K. Chakrabarti,  Infinite-range Ising ferromagnet in a
    time-dependent transverse magnetic field: quench and ac dynamics
    near the quantum critical point. Phys. Rev. B \textbf{74}, 144423 (2006).}

\bibitem{Milburn97}{G. J. Milburn, J. Corney, E. M.
    Wright,  and D. F. Walls,  Quantum dynamics of an atomic
    Bose-Einstein condensate in a double-well potential. Phys. Rev. A \textbf{55}, 4318 (1997).}


\bibitem{albiez05}{M. Albiez, R. Gati, J. F\"{o}lling, S. Hunsmann, 
    M. Cristiani, and M. K. Oberthaler, Direct observation of
    tunneling and nonlinear self-trapping in a single bosonic
    Josephson junction. Phys. Rev. Lett. \textbf{95},
    010402 (2005).}
    
\bibitem{Levy07}{S. Levy, E. Lahoud, I. Shomroni, and
    J. Steinhauer, The a.c. and d.c. Josephson effects in a
    Bose-Einstein condensate. Nature \textbf{449}, 579 (2007).}
    
\bibitem{zibold10}{T. Zibold, E. Nicklas, C. Gross,  and M. K.
    Oberthaler,  Classicial bifurcation at the Transition from
    Rabi to Josephson dynamics. Phys. Rev. Lett. \textbf{105}, 204101 (2010).}    
    
\bibitem{Gerving12}{C.S. Gerving, T. M. Hoang, B.J. Land, M. Anquez, C.D. Hamley, and M.S. Chapman, Non-equilibrium dynamics of an unstable quantum pendulum explored in a spin-1 Bose-Einstein condensate. Nat. Commun. \textbf{3} 1169 (2012).}    

\bibitem{Valtolina15}{G. Valtolina, A. Burchianti, A. Amico, E. Neri, K. Xhani, J. A. Seman, A. Trombettoni, A. Smerzi, M. Zaccanti, M. Inguscio, and G. Roati, Josephson effect in fermionic superfluids across the BEC-BCS crossover. Science \textbf{350}, 1505 (2015).}

\bibitem{trenkwalder16}{A. Trenkwalder, G. Spagnolli, 
    G. Semeghini, S. Coop, M. Landini, P. Castilho,
    L. Pezz\`{e}, G. Modugno, M. Inguscio, A. Smerzi, and M. Fattori, Quantum phase transitions with parity-symmetry breaking and hysteresis. Nat. Phys. \textbf{12}, 826 (2016).} 


\bibitem{Rigas13}{I. Rigas, A. B. Klimov, L. L. S\'{a}nchez-Soto. and G. Leuchs, New Journal of Physics \textbf{15},  043038 (2013).}

\bibitem{luis02}{A. Luis, Degree of polarization in quantum optics. Phys. Rev. A \textbf{66}, 013806 (2002).}

\bibitem{korolkova05}{N. Korolkova and R. Loudon,  Nonseparability and
    squeezing of continuous polarization variables. Phys. Rev. A \textbf{71}, 032343 (2005).}

\bibitem{Leggett}{A. J. Leggett  \textit{Chance and Matter} (Les Houches 1986, Session XLVI) ed J Souletie et al (North-Holland, Amsterdam, 1987).}

\bibitem{Nieto93}{M. M. Nieto, Quantum Phase and Quantum Phase Operators: Some Physics and Some History. Physica Scripta. \textbf{T48}, 5 (1993), and references therein.}

\bibitem{Polkovnikov10}{A. Polkovnikov, Phase space representation of quantum dynamics. Annals of Phys. \textbf{325} 1790 (2010).}

\bibitem{Javan13}{J. Javanainen and J.  Ruostekoski, Emergent
    classicality in continuous quantum measurements. New J. Phys. \textbf{15}, 013005 (2013).}


\bibitem{smerzi97}{A. Smerzi, S. Fantoni, S. Giovanazzi, and S. R. Shenoy,  Quantum coherent atomic tunneling between two trapped Bose-Einstein condensates. Phys. Rev. Lett. \textbf{79}, 4950 (1997).}
        
\bibitem{Veksler15}{H. Veksler and S. Fishman, Semiclassical analysis of Bose-Hubbard dynamics. New J. Phys. \textbf{17}, 053030 (2015).}    
    
\bibitem{Berry66}{M.V. Berry, \textit{The Diffraction of Light by Ultrasound}
(Academic, New York, 1966).}    

\bibitem{Leibscher04}{M. Leibscher, I. Sh. Averbukh, P. Rozmej, and R. Arvieu, Phys. Rev. A  \textbf{69}, 032102 (2004).}

\bibitem{Huckans09}{J. H. Huckans, I. B. Spielman, B. L. Tolra, W. D. Phillips,
and J. V. Porto, Quantum and classical dynamics of a Bose-Einstein condensate in a large-period optical lattice. Phys. Rev. A \textbf{80}, 043609 (2009).}

\bibitem{Desyatnikov11}{A. S. Desyatnikov, M. R. Dennis, and A. Ferrando, All-optical discrete vortex switch. Phys. Rev. A \textbf{83}, 063822 (2011).}

\bibitem{haake09}{F. Haake, \textit{Quantum Signatures of Chaos},
    $3^{\mathrm{rd}}$ edition (Springer, Berlin, 2009).}
    
\bibitem{pegg89}{D. T. Pegg  and S. M. Barnett,  Phase properties of the quantized single-mode electromagnetic field. \textit{Phys. Rev. A}
    \textbf{39}, 1665 (1989).}       

\bibitem{Poisson}{The Poisson resummation formula is $\sum_{m=-\infty}^{\infty}g(m)=\sum_{k=-\infty}^{\infty} \int_{-\infty}^{\infty}g(m) \ \mathrm{e}^{-2 \pi \mathrm{i} m k} \ \mathrm{d} m$ and is exact. This formula requires $m$ to take values in the range $-\infty \le m \le \infty$ and this restricts its applicability in our case to the semiclassical regime $N \gg 1$. See \cite{SM} for details.}

\bibitem{Fujikawa79}{K. Fujikawa, Path-Integral Measure for Gauge-Invariant Fermion Theories. Phys. Rev. Lett. \textbf{42}, 1195 (1979).}

\bibitem{Esteve02}{J. G. Esteve, Origin of the anomalies: The modified Heisenberg equation. Phys. Rev. D \textbf{66}, 125013 (2002).}

\bibitem{Coon02}{S. A. Coon and B. R. Holstein, Anomalies in quantum mechanics: The $1/r^2$ potential. Am. J. Phys. \textbf{70}, 513 (2002).} 


\bibitem{Olshanii10}{M. Olshanii, H. Perrin, and V. Lorent, Example of a Quantum Anomaly in the Physics of Ultracold Gases, Phys. Rev. Lett. \textbf{105}, 095302 (2010).}

\bibitem{Heisenbergalgebra}{In other words, the algebra opens from a Heisenberg to an su(2) algebra.}


\bibitem{Harty14}{T. P. Harty, D. T. C. Allcock, C. J. Ballance, L. Guidoni, H. A. Janacek, N. M. Linke, D. N. Stacey, and D. M. Lucas, High-Fidelity Preparation, Gates, Memory, and Readout of a Trapped-Ion Quantum Bit, Phys. Rev. Lett. \textbf{113}, 220501 (2014).}  

\bibitem{nicklas15}{E. Nicklas, M. Karl, M. H\"{o}fer, A. Johnson, W. Muessel, H. Strobel, J. Tomkovi\v{c}, T. Gasenzer, and M. K. Oberthaler, Observation of Scaling in the Dynamics of a Strongly Quenched Quantum Gas, Phys. Rev. Lett. \textbf{115}, 245301 (2015).}

\bibitem{Kirkby17}{W. Kirkby, J. Mumford and D. H. J. O'Dell,  arXiv:1710.01289}

\end{thebibliography}
\end{document}